\documentclass[a4paper,12pt]{article}
\usepackage{amsmath,amsfonts,amssymb,cite}
\usepackage[dvips]{graphicx}

\renewcommand{\Im}{\text{Im\,}}

\newcommand{\als}{\alpha_s}
\newcommand{\qqb}{\langle\bar{q}q\rangle}

\newcommand{\mqq}{\langle m_q\bar{q}q\rangle}

\newcommand{\gmv}{\left\langle\frac{\als}{\pi}\left(G_{\mu\nu}^a\right) ^2\right\rangle}

\newcommand{\lsb}{\left[}
\newcommand{\rsb}{\right]}

\begin{document}

\begin{center}
{\Large\bf Electromagnetic couplings of radially excited light vector mesons from QCD sum rules}
\end{center}
\bigskip
\begin{center}
{\large S. S. Afonin
and T. D. Solomko}
\end{center}

\begin{center}
{\it Saint Petersburg State University, 7/9 Universitetskaya nab.,
St.Petersburg, 199034, Russia}
\end{center}

\begin{abstract}
The couplings of unflavored vector mesons to the $e^+e^-$ annihilation are an important
source of information on the nature and structure of these resonances which play a
prominent role in the hadron phenomenology. The couplings of many radially excited
heavy vector mesons are measured, while the corresponding couplings for light vector mesons
are not known. We propose a phenomenological method allowing to estimate these unknown couplings.
The method is suggested by our observation that the electromagnetic coupling of the $n$-th radial excitation
of $S$-wave heavy vector meson decouples from the $e^+e^-$ annihilation with nearly
exponential rate with $n$. It becomes natural to assume that the same effect takes place
in the light vector mesons and this would allow to estimate the unknown couplings.
We tested this assumption with the help of a generalized version of borelized QCD spectral
sum rules saturated by a linear radial trajectory of meson states taken from the
phenomenology. This leads to a consistent setup which is able to predict the
decoupling rate. The calculated rate turned out to be almost the same as in the heavy
vector mesons. Our result may be interpreted as an effective account for non-trivial
electromagnetic formfactor in meson decays to $e^+e^-$, thus looking beyond the large-$N_c$ limit.
On the one hand, we argue that the given decoupling does not necessary contradict
to the previous large-$N_c$ results for the Regge like meson spectra.
\end{abstract}

\section{Introduction}

The Neutral Vector Mesons (NVMs) play a very important role in the hadron physics
due to their direct coupling to photons. As is well known the high-energy photons
interact with hadrons mainly by means of conversion into a NVM while the contribution of
straight interaction with hadron electric charge is almost negligible. This gave rise
to the famous hypothesis of vector meson
dominance~\cite{vmd1,vmd2_1,vmd2_2,vmd2_3}.
In contrast to gapless photons the NVMs possess a discrete spectrum of radial excitations.
The couplings to the $e^+e^-$ annihilation of radially excited heavy NVMs are known from the
measured electromagnetic decay widths~\cite{pdg}. The same couplings for the light NVMs, however,
are not known except the ground $\rho^0$, $\omega$, and $\phi$ states. This uncertainty
may cause serious problems for checking various theoretical models, especially for models
formulated in the large-$N_c$ limit of QCD~\cite{hoof,witten} where all meson resonances
represent narrow and hence well defined states.

We should briefly explain the point. Some time ago it was quite popular to address the
spectroscopy of light mesons via the large-$N_c$ extensions of QCD spectral sum rules
(see, e.g.,
Refs.~\cite{sr_sh1,sr_sh2,duality1_1,duality1_2,duality1_3,sr0_1,sr0_2,sr0_3,sr0_4,sr0_5,sr0_6,sr1,sr2,sr3,sr4,sr5,sr6,sr7,sr8,sr9,sr10,sr11,sr12,sr13,sr14,sr15,sr16,sr17,sr18}).
Let us consider the $\rho$-meson as an example.
The basic theoretical object in this approach is the two-point correlator
$\Pi(Q^2)$ defined by
\begin{equation}
\label{1}
\Pi_{\mu\nu}=(q_\mu q_\nu-g_{\mu\nu}q^2)\Pi(Q^2),
\end{equation}
where $\Pi_{\mu\nu}$ represents the T-product of two vector
currents interpolating the neutral $\rho^0$-meson,
\begin{equation}
\label{2}
\Pi_{\mu\nu}=i\int d^4x\, e^{iqx}\left\langle0|\text{T}\left\{j_\mu(x),j_\nu(0)\right\}\right|0\rangle,
\end{equation}
$$j_\mu=\frac12(\bar{u}\gamma_\mu u-\bar{d}\gamma_\mu d).$$
Here $u$ and $d$ are quark fields, and $q$ is the photon space-like
momentum, $q^2=-Q^2$. In the large-$N_c$ limit, the correlator is saturated by one-hadron states,
\begin{equation}
\label{3}
\Pi(q^2)=\sum_n\frac{F_n^2}{q^2-m_n^2+i\epsilon}.
\end{equation}
The resonances are regarded as infinitely narrow because the large-$N_c$ scaling of meson masses
is $m_n\sim N_c^0$ while the scaling of full meson width is $\Gamma_n\sim m_n/N_c$,
so $\Gamma_n\rightarrow 0$ in the limit $N_c\rightarrow\infty$. The quantities $F_n$ in
residues of relation~\eqref{3} represent electromagnetic couplings defined by
\begin{equation}
\label{4}
\langle \text{vac}|j^\mu|\rho_n\rangle=\frac12 eF_nm_n\varepsilon^\mu_n,
\end{equation}
where $\varepsilon^\mu_n$ is the polarization vector of the vector meson $\rho_n$.
At large $q^2$ the leading perturbative contribution to $\Pi(q^2)$ is logarithmic,
\begin{equation}
\label{5}
\Pi(q^2)\sim\log{q^2}.
\end{equation}
It is clear that the consistency of Eq.~\eqref{3} with Eq.~\eqref{5} can be achieved only if
infinite number of states are included in Eq.~\eqref{3}~\cite{witten}.

For further analysis one needs to postulate the form of meson spectrum. Basing on various
theoretical (starting from the Veneziano model, for more recent suggestions see, e.g.,
Refs.~\cite{linear1,linear2,linear3,linear4,linear5,linear6,linear7,linear8,linear9,linear10,linear11,linear12,linear13,linear14,linear15,linear16})
and phenomenological (see, e.g.,
Refs.~\cite{phen,phen2,phen3_0,phen3_1,phen3_2,phen3_3,phen3_4,phen3_5,phen3_6,phen3_7,phen3_8,phen4_0,phen4_1})
expectations the spectrum was usually interpolated by a
simple linear Regge-like trajectory of radial states,
\begin{equation}
\label{6}
m_n^2 = a n + m_0^2, \qquad n=0,1,2,\dots,
\end{equation}
sometimes with some non-linear corrections.

In the case of linear ansatz~\eqref{6}, the logarithmic asymptotic~\eqref{5} can be obtained
only if $F_n^2$ in Eq.~\eqref{3} behave with $n$ as
\begin{equation}
\label{7}
F_n^2\sim\frac{dm_n^2}{dn}\sim\text{const},
\end{equation}
at least for $n\rightarrow\infty$.
In this way one arrives at the standard large-$N_c$ prediction for the e/m constants $F_n$:
They are expected to be almost $n$-independent for large enough $n$,
i.e. all highly excited states should couple to the $e^+e^-$ annihilation
nearly equally. The problem is that we still cannot check this prediction --- the e/m constants
$F_n$ are related to the e/m decay widths,
\begin{equation}
\label{8}
\Gamma_{\rho_n\rightarrow e^+e^-}=\frac{4\pi\alpha^2F_n^2}{3m_n},
\end{equation}
which have not been reliably measured for the radially excited states, $n>0$, in the light quark
sector. A pertinent question of validity of relation~\eqref{7} emerges also in the soft-wall
holographic model of QCD~\cite{son2} and in numerous extensions of this model (see, e.g., discussions
in Refs.~\cite{Brodsky1,Brodsky2,Brodsky3,Brodsky4,Brodsky5}). The given bottom-up holographic approach is congenial with the method of large-$N_c$ QCD
sum rules~\cite{holog1,holog2} and the both are based on at least asymptotic validity of Eq.~\eqref{7}.
In the situation of absence of a direct experimental information on important couplings $F_n$, may
we test the prediction~\eqref{7} somehow indirectly?

In the present work, we will give arguments disfavoring the validity of Eq.~\eqref{7} in the real world
with $N_c=3$ if the relation~\eqref{8} holds.

The paper is organized as follows. In the introductory Section~1, we have formulated the problem.
In Section~2, we recall briefly how the relation~\eqref{7} was originally proposed and revise the
original estimates. A comparison with the situation in heavy vector mesons is made in Section~3.
Motivated by this comparison we propose in Section~4 an ansatz for $F_n^2$ and perform a test for
this ansatz in the framework of QCD spectral sum rules. Concluding discussions on possible
interpretations of our result are given in the final Section~5.

\section{Extended Vector Meson Dominance}

We find instructive to begin with a brief reminder of an old history of appearing the relation~\eqref{7} from the
hypothesis of extended Vector Meson Dominance (VMD). The classical VMD hypothesis put forward by Sakurai~\cite{sakurai}
states that the vector current in~\eqref{2} can be replaced by the field of vector meson,
\begin{equation}
\label{9}
j^\mu=e\frac{m_\rho^2}{2f_\rho}\rho^\mu.
\end{equation}
The full vector current includes of course the field of $\omega$ and $\phi$ meson but this will be not
essential for our purposes. The dimensionless coupling $f_\rho$ in~\eqref{9} is related to $F_\rho$ in~\eqref{4} as
$$
f_\rho=\frac{m_\rho}{F_\rho}.
$$
We prefer to use temporary this traditional for the VMD hypothesis notation.

At the beginning of 1970s
it was realized that the VMD works well at low enough energies and at higher energies the predictions
can be improved if the second $\rho$-meson, the excited $\rho'$ one, is added to~\eqref{9}. This resonance
was referred to as $\rho(1600)$~\cite{bramon2}. It was looking natural to include into the identity~\eqref{9} further
higher mass vector mesons up to infinite number. In particular, the authors of Ref.~\cite{bramon} (and independently
Sakurai~\cite{sakurai2}) argued that the infinite number of states in~\eqref{9} is able to describe the scaling of
the inelastic structure functions. This alternative scheme to parton models has subsequently been interpreted as a
particular manifestation of the quark-hadron duality. In this model, the total $e^+e^-$ annihilation cross
section into hadrons is given by
\begin{equation}
\label{10}
\sigma_{e\bar{e}\rightarrow\text{h}}(s)=\frac{4\pi^2\alpha^2}{s^\frac32}\sum_n\frac{m_n^3}{f_n^2}\frac{m_n\Gamma_n}{(s-m_n^2)^2+m_n^2\Gamma_n^2}.
\end{equation}
If asymptotically $\sigma_{e\bar{e}\rightarrow\text{h}}(s)$ scales as $1/s$ then the sum in~\eqref{10}
must behave as a function of c.m. energy $s$ as $\sqrt{s}\times\text{const}\times\Theta(s-m_0^2)$
implying for large $n$
\begin{equation}
\label{11}
\frac{m_n^2}{f_n^2}\rightarrow\text{const},
\end{equation}
which is the relation~\eqref{7}. The analysis of existed data in terms of a broad $\rho(1600)$ which was
performed in Ref.~\cite{bramon2} resulted in the estimate $f_{\rho'}/f_\rho\approx\sqrt{5}$ (see below) suggesting that
the relation~\eqref{11} is approximately satisfied already for $\rho(1600)$. Indeed, we have from these
numbers $(m_{\rho}/f_{\rho})^2/(m_{\rho'}/f_{\rho'})^2\approx1.2$. From this observation the authors of Ref.~\cite{bramon}
assumed for simplicity that~\eqref{11} is true for all $n$.

Let us now revise the analysis of Ref.~\cite{bramon2} using the modern data. The partial width of the
$\rho'\rightarrow\omega\pi\rightarrow4\pi$ decay mode was predicted to be
\begin{equation}
\label{12}
\Gamma_{\rho'\rightarrow\omega\pi}=2.4\left(\frac{f_{\rho'}}{f_{\rho}}\right)^2\Gamma_{\omega\rightarrow3\pi}.
\end{equation}
Making use of $SU(3)$ and phase-space evaluation techniques and substituting
$\Gamma_{\omega\rightarrow3\pi}\approx10$~MeV the authors of Ref.~\cite{bramon2} obtained for the main
$\rho'\rightarrow\text{VP}$ decay modes (in MeV)
\begin{equation}
\label{13}
\Gamma_{\rho'\rightarrow\omega\pi}=24\left(\frac{f_{\rho'}}{f_{\rho}}\right)^2, \quad
\Gamma_{\rho'\rightarrow\rho\eta}=7\left(\frac{f_{\rho'}}{f_{\rho}}\right)^2, \quad
\Gamma_{\rho'\rightarrow K^*K}=4\left(\frac{f_{\rho'}}{f_{\rho}}\right)^2.
\end{equation}
The pseudoscalar-pseudoscalar decay modes were estimated to be equal to
$\Gamma_{\rho'\rightarrow\text{PP}}\simeq8\left(\frac{f_{\rho'}}{f_{\rho}}\right)^2$~MeV.
By adding all these rates they obtained
$\Gamma_{\rho'\rightarrow \text{VP,PP}}\simeq43\left(\frac{f_{\rho'}}{f_{\rho}}\right)^2$~MeV.
Using $\Gamma_{\rho'}\simeq215$~MeV the final result was $(f_{\rho'}/f_\rho)^2\approx5$.

The modern value for the width of decay $\omega\rightarrow3\pi$ is
$\Gamma_{\omega\rightarrow3\pi}\approx7.57$~MeV~\cite{pdg}. Thus the numbers in~\eqref{13}
should be multiplied by the factor of 0.757 and this leads to
$\Gamma_{\rho'}\simeq34\left(\frac{f_{\rho'}}{f_{\rho}}\right)^2$~MeV.

Since 1988 the old resonance $\rho(1600)$ has been replaced in PDG by two resonances
$\rho(1450)$ and $\rho(1700)$~\cite{pdg}. The former is usually interpreted as the first radial
$S$-wave excitation of $\rho(770)$. The substitution of $\Gamma_{\rho(1450)}=400\pm60$~MeV
and $m_{\rho(1450)}=1465\pm25$~MeV~\cite{pdg} leads to the estimate
$(f_{\rho'}/f_\rho)^2\approx11.6\pm1.7$. With the modern data we hence obtain
$(m_{\rho}/f_{\rho})^2/(m_{\rho'}/f_{\rho'})^2\approx3.2(3)$ instead of 1.2 above.
Thus the relation~\eqref{11} is in a sharp contradiction for $\rho'$. Rephrasing this
in terms of couplings $F_n$ in~\eqref{7} we have the estimate,
$$
\beta\equiv\frac{F_1^2}{F_0^2}\approx0.31(3).
$$

It is interesting to compare the given value of $\beta$ with other estimates
existing in the literature. We found some estimates for $F_1$ (i.e. for $F_{\rho'}$)
which lead to the following predictions: $\beta\approx0.3$~\cite{henner} (from a model of $S$-matrix
unitarity for overlapping resonances); $\beta\gtrsim0.3$~\cite{YP} (from contribution of $\rho'$
to the electromagnetic pion mass difference); $\beta\approx0.25$~\cite{nedelko}
(from a specific model of gluon vacuum); $\beta\approx0.27-0.46$~\cite{bakker}
(from a specific relativized potential quark model); $\beta\approx0.59(12)$~\cite{Jiang}
(from finite energy QCD sum rules saturated by two resonances). It should be mentioned that
the last estimate obtained with the help of the least square fitting method is consistent
with the estimates which we will get in the present paper.

\section{A hint from the heavy quark sector}

The subject discussed in the previous Section is almost 50 years old. However we still
do not have any significant progress in getting reliable experimental data on e/m couplings
of light excited NVMs. Such a situation looks unfortunate to say the least. On the other hand,
a serious progress was achieved in the heavy NVMs. The vector charmonia and bottomonia represent
direct analogues of light $\rho^0$, $\omega$, and $\phi$ mesons and share common properties.
It is worth to mention a recent observation that even their radial Regge trajectories look
very similar if one subtracts the quark masses $m_q$ from the corresponding masses of heavy
mesons: $(m_n-2m_q)^2=an+m_0^2$, where the slope $a$ and intercept $m_0^2$ are nearly universal
for both light and heavy mesons~\cite{prd1,prd2,prd3,prd4}. It looks natural to expect that their electromagnetic
properties are also similar.

The $e^+e^-$ decay width of many excited $\psi$ and $\Upsilon$ mesons was measured with a
satisfactory accuracy~\cite{pdg}. By extracting the e/m couplings $F_n$ from these data with
the help of relation~\eqref{8} we observe that the prediction~\eqref{7} is not fulfilled.
One has instead a nearly exponential decrease with $n$. To quantify this observation we
will interpolate the given decrease in the form of geometrical progression,
\begin{equation}
\label{14}
\beta^n\equiv\frac{F_n^2}{F_0^2}=\frac{\Gamma_{e\bar{e},n}m_n}{\Gamma_{e\bar{e},0}m_0},
\end{equation}
where $n=0$ refers to the ground state. In Table~1 we show the extracted values of $\beta$
for those vector mesons which are supposed to be the $S$-wave radial excitations of
$J/\psi(1S)$ and $\Upsilon(1S)$ states.

\begin{table}[h!]
\caption{\small The masses, $e^+e^-$ decay widths and values of $\beta$ for the $S$-wave
heavy vector mesons.}
\bigskip
$\begin{array}{|c||c|c|c|c|c|}
 \hline
  & J/\psi(1S) & \psi(2S) & \psi(4040) & \psi(4160) & \psi(4415) \\
 \hline
 \hline
 \text{m, MeV} & 3097 & 3686 & 4039\pm1 & 4191\pm5 & 4421\pm4 \\
 \hline
 \Gamma_{e\bar{e}},\,\text{keV} & 5.55\pm0.14 & 2.34\pm0.04 & 0.86\pm0.07 & 0.48\pm0.22 & 0.58\pm0.07 \\
 \hline
 \beta & \text{---} & 0.50\pm0.02 & 0.45\pm0.03 & 0.49\pm0.08 & 0.62\pm0.02 \\
 \hline
\end{array}$
\end{table}
\vspace{-0.5cm}
\begin{table}[h!]
$\begin{array}{|c||c|c|c|c|c|}
 \hline
  & \Upsilon(1S) & \Upsilon(2S) & \Upsilon(3S) & \Upsilon(4S) & \Upsilon(11020) \\
 \hline
 \hline
 \text{m, MeV} & 9460 & 10023 & 10355 & 10579\pm1 & 10988^{+11}_{-3} \\
 \hline
 \Gamma_{e\bar{e}},\,\text{keV} & 1.34\pm0.02 & 0.61\pm0.01 & 0.44\pm0.01 & 0.27\pm0.03 & 0.13\pm 0.03 \\
 \hline
 \beta & \text{---} & 0.48\pm0.01 & 0.60\pm0.01 & 0.61\pm0.02 & 0.58\pm0.04 \\
 \hline
\end{array}$
\end{table}

It is well seen that the values of $\beta$ are rather stable and concentrate near
$\beta\approx0.5-0.6$. This property is at odds with the large-$N_c$ prediction
$\beta\approx1$ in~\eqref{7}. It is tempting to assume that in the light quark sector
the situation can be similar. The given assumption, in principle, can be tested
within the framework of QCD spectral sum rules. In the next Section, we propose such a test.

\section{Electromagnetic couplings from QCD sum rules}

Let us substitute the ansatz~\eqref{14} into the classical SVZ sum rules~\cite{svz} extended to
the case of arbitrary number of resonances (often called "the finite energy sum rules"). The
method of QCD sum rules is based on the Operator Product Expansion (OPE) of the two-point
correlator~\eqref{2} in Euclidean space,
\begin{multline}
\label{15}
\Pi(Q^2)=\frac{1}{8\pi^2}\left[\left(1+\frac{\als}{\pi}\right)\ln\frac{\mu^2}{Q^2} - \frac{6m_q^2}{Q^2}\right] \\
+ \frac{\mqq}{Q^4} + \frac{1}{24 Q^4}\gmv -\frac{14}{9}\frac{\pi\als}{Q^6}\qqb^2,
\end{multline}
where $q$ stands for $u$ or $d$ quark, the coefficient in front of
the last term is given in the large-$N_c$ limit, and further $\mathcal{O}(Q^{-8})$ terms
are neglected. The non-perturbative power corrections in the second line are given in form of
various vacuum condensates. To improve the convergence of the OPE and increase the relative
contribution of the ground state one applies the Borel transform,
\begin{equation}
\label{16}
L_M\Pi(Q^2)=\lim_{\substack{Q^2,n\rightarrow\infty\\Q^2/n=M^2}}\frac{1}{(n-1)!}(Q^2)^n\left(-\frac{d}{dQ^2}\right)^n\Pi(Q^2),
\end{equation}
to the OPE~\eqref{15} and gets
\begin{multline}
\label{17}
L_M\Pi(Q^2)=\frac{1}{8\pi^2}\left(1+\frac{\als}{\pi}-\frac{6m_q^2}{M^2}\right) \\ + \frac{\mqq}{M^4} +\frac{1}{24 M^4}\gmv -\frac{7}{9}\frac{\pi\als}{M^6}\qqb^2.
\end{multline}

The vector correlator $\Pi(Q^2)$ satisfies a dispersion relation
with one subtraction,
\begin{equation}
\label{18}
\Pi(q^2)=\frac{1}{\pi}\int_{4m_q^2}^\infty ds \frac{\Im\Pi(s)}{s-q^2 +i\varepsilon}+\Pi(0).
\end{equation}
In the classical SVZ sum rules, the correlator~\eqref{18} is saturated by one resonance
plus perturbative continuum. We will saturate by arbitrary number of states with the linear
spectrum~\eqref{6}. The decay width in this method is neglected (aside from the large-$N_c$ arguments,
this approximation can be motivated phenomenologically by the observation that on average the full width scales as
$\Gamma_n\simeq0.1m_n$~\cite{phen4_0,phen4_1} so one expects an accuracy on the level of 10\%). We assume that the contribution of
higher mass states is rapidly decreasing. This allows not to bother about an exact energy cutoff and
use instead an infinite number of states. We will see the consistency of this assumption
{\it aposteriori}.

In the case of ansatz~\eqref{14}, the imaginary part of $\Pi(Q^2)$ in the
resonance representation~\eqref{3} takes the form
\begin{equation}
\label{19}
\Im\Pi(q^2)=\sum_{n}\pi\beta^n F_0^2 \delta(q^2-m_n^2).
\end{equation}
The Borel transform of~\eqref{18} is~\cite{svz}:
\begin{equation}
\label{20}
L_M \Pi(Q^2) = \frac{1}{\pi M^2}\int_{0}^\infty e^{-s/M^2}\Im \Pi(s)ds
= \frac{F_0^2}{M^2}\sum_{n}\beta^n e^{-m_n^2/M^2},
\end{equation}
where we neglected the $\mathcal{O}(m_q^2)$ contribution.
Substituting the linear spectrum~\eqref{6} and summing the contributions
from infinite number of states we obtain
\begin{equation}
\label{21}
L_M \Pi(Q^2) = \frac{F_0^2}{M^2}\frac{e^{-m_0^2/M^2}}{1-\beta e^{-a/M^2}}.
\end{equation}

The first sum rule follows from equating the relations~\eqref{15}
and~\eqref{21},
\begin{multline}
\label{22}
\frac{F_0^2  e^{-m_0^2/M^2}}{1-\beta e^{-a/M^2}} =
\frac{M^2}{8\pi^2} \lsb
    1 + \frac{\als}{\pi} - \frac{6m_q^2}{M^2} \right.\\
\left.   + \frac{8\pi^2}{M^4}\mqq +
    \frac{\pi^2}{3 M^4}\gmv -
    \frac{56}{9}\frac{\pi^3 \als}{M^6}\qqb^2
\rsb.
\end{multline}
The second sum rule arises after taking derivative of Eq.~\eqref{22} with
respect to $1/M^2$~\cite{svz}. The mass squared of the ground state appears
directly in the fraction $-\frac{d(22)}{d(1/M^2)}/(22)$,
\begin{equation}
\label{23}
m_0^2=M^2\frac{h_0 - \frac{h_2}{M^4} - \frac{2h_3}{M^6}}{h_0 + \frac{h_1}{M^2} + \frac{h_2}{M^4} + \frac{h_3}{M^6}}
-\frac{a}{1-\beta e^{-a/M^2}},
\end{equation}
where the condensate terms $h_i$ are presented in Table~2.

\begin{table}[ht]
\begin{center}
\caption{\small The condensate contributions $h_i$ and their numerical values used in our work (see text).}
\bigskip
$\begin{array}{|c|c|c|c|c|}
 \hline
   & h_0 & h_1 & h_2 & h_3 \\
 \hline
  \text{Theor.} & 1 + \frac{\als}{\pi} & -6m_q^2 & 8\pi^2\mqq + \frac{\pi^2}{3}\gmv & -\frac{56}{9}\pi^3\als\qqb^2 \\
 \hline
  \text{Numer.} & 1 & 0 & 0.032 & -0.030 \\
 \hline
\end{array}$
\end{center}
\end{table}

The first term in the r.h.s. of Eq.~\eqref{23} corresponds to the
limit $s_0\rightarrow\infty$ in the canonical expressions for the
meson masses in the SVZ sum rules~\cite{svz}. The energy cutoff
$s_0$ is infinite in our case as we take into account the infinite
number of radial excitations. The second term reflects
contribution of highly excited states ($a\neq0$) with,
generally speaking, decreasing residues in the corresponding poles
($0<\beta\leq1$).

We will set $h_0=1$ since taking the perturbative threshold
$s_0\rightarrow\infty$ (infinite number of radial states) we
should have  $\als\rightarrow0$ due to the asymptotic
freedom. This is consistent with the Borel transform~\eqref{16}.
Indeed, the coupling $\als$ is a function of the momentum
injected --- the corresponding one-loop expression in Euclidean space
is $\als\sim1/\ln\frac{Q^2}{\Lambda_{\text{QCD}}^2}$.
In finite energy sum rules, one neglects this slow
running of $\als$ with $Q^2$ setting $\mu^2 = s_0$. But in the
infinite energy region this is not a good approximation and the
running of $\als$ must be taken into account.
The induced correction to the unit operator in OPE will be then
proportional to
$\ln\frac{\mu^2}{Q^2}/\ln\frac{Q^2}{\Lambda_{\text{QCD}}^2}$.
One can see that this term becomes zero after applying the Borel
transform~\eqref{16} --- its leading part is proportional to
$\lim_{Q^2\rightarrow\infty}\ln\frac{\mu^2}{\Lambda_{\text{QCD}}^2}/
\ln^2\frac{Q^2}{\Lambda_{\text{QCD}}^2}$.
For the same reason other loop corrections to the unit
operator in the OPE disappear after the Borel transform.

Now we should fix the other input parameters.
The values of gluon and quark condensates are taken from the work~\cite{svz}:
$\gmv=(330\,\text{MeV})^4,$ and
$\qqb=-(250\,\text{MeV})^3$. The first value is scale-independent
while the second one is taken roughly at the scale $\mu=1$~GeV.
From the Gell-Mann--Oakes --Renner relation,
$m_\pi^2f_\pi^2=-(m_u+m_d)\qqb$, with the pion mass
$m_\pi=140$~MeV~\cite{pdg} and pion weak decay constant
$f_\pi=92.4\,\text{MeV}$, one gets $m_u+m_d\approx
10.7\,\text{MeV}\Bigr|_{\mu=1\,\text{GeV}}$. We consider the
isospin limit for the masses of current quarks, $m_u=m_d\equiv
m_q$. Thus we get a numerical value for another renormalization
invariant condensate of dimension four, $\mqq=-(95.6\,\text{MeV})^4$.
The scale-dependent term $\mathcal{O}(m_q^2)$ is numerically very small
and should be neglected within our accuracy.
All these inputs lead to the values of $h_1$ and $h_2$ in Table~2.
The operator $\als(\bar{q}q)$ has a small anomalous dimension. We
will regard the corresponding v.e.v. $\als\qqb$ as a constant. The
numerical value for $h_3$ in Table~2 is taken from Ref.~\cite{svz}.

The last free parameter to be fixed is the slope $a$ in the linear
spectrum~\eqref{6}. The fixation of slope from the data on excited
$\rho$-mesons is somewhat ambiguous as these data admit various
interpretations~\cite{pdg}. We will stick with a conservative point
and take a value averaged over many linear trajectories~\cite{phen2},
$a=1.14\,\text{GeV}^2$.

\begin{figure}[ht]
    \includegraphics[scale=1]{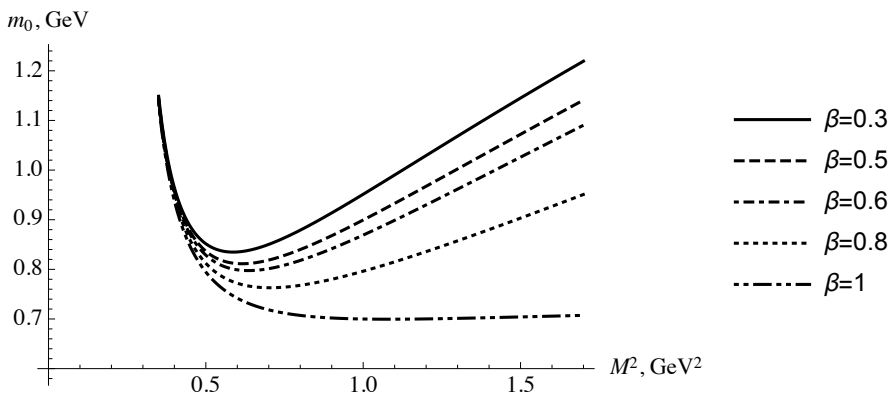}
    \caption{\small The mass of $\rho$ meson on the Borel plane at different values of $\beta$.}
\end{figure}
\begin{figure}[ht]
    \includegraphics[scale=1]{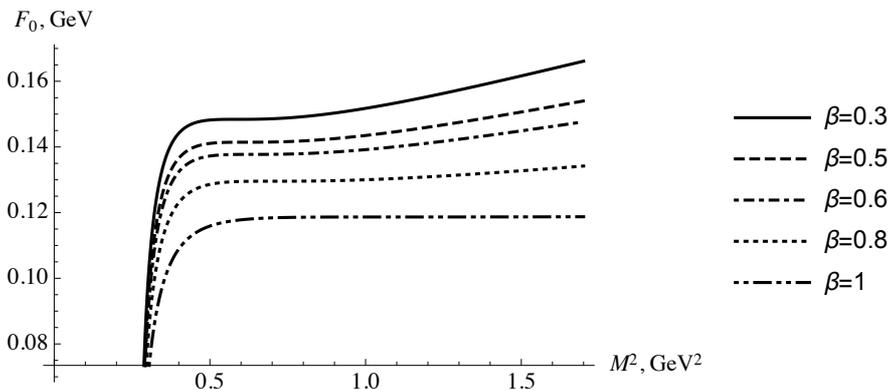}
    \caption{\small The e/m constant of $\rho$ meson on the Borel plane at different values of $\beta$.}
\end{figure}

The behavior of ground state mass $m_0$ calculated from Eq.~\eqref{23}
and of e/m coupling $F_0$ from Eq.~\eqref{22} as a function of Borel
parameter $M^2$ at different values of $\beta$ is displayed on
Fig.~1 and Fig.~2 respectively. The existence of stability region ---
the so-called "Borel window" --- is well seen for $0.5<M^2<0.8\,\text{GeV}^2$ when $\beta<1$.
The predictions of SVZ sum rule method refer to that stability region only.
It should be remarked that we used the narrow-width approximation
where the values of $m_0$ and $F_0$ may be different from the
experimental ones, $m_\rho=775$~MeV and $F_\rho=156(2)$~MeV~\cite{pdg}.
For instance, the unitarized chiral perturbation theory predicts
the enhancement of $m_\rho$ by 40-60~MeV when taking the zero-width
limit~\cite{unitar1,unitar2,unitar3}. On the other hand, the Regge phenomenology of light and
heavy vector mesons suggests that their ground states lie always below
the corresponding radial linear trajectories~\cite{prd1,prd2,prd3,prd4}, i.e. the linear
ansatz~\eqref{6} should predict a somewhat enhanced value for $m_0$.
Since we made use of both the zero-width approximation and the linear radial trajectories,
we should reproduce the value of $m_0$ expected in these approximations,
i.e., roughly speaking, we should use $m_\rho=800-830$~MeV as a reference value.
The given mass is achieved near $\beta\approx0.5-0.6$ (see Fig.~1). Exactly
this interval of $\beta$ was observed for the heavy vector mesons in the
previous Section! This is our main result.


\begin{figure}[ht]
    \includegraphics[scale=1]{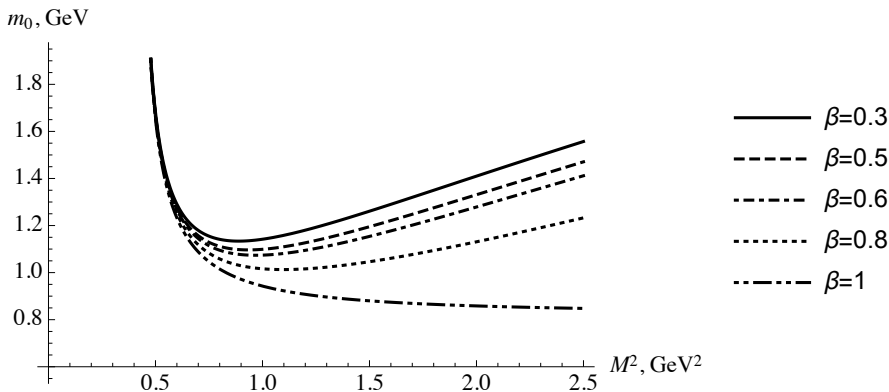}
    \caption{\small The mass of $\phi$ meson on the Borel plane at different values of $\beta$.}
\end{figure}
\begin{figure}[ht]
    \includegraphics[scale=1]{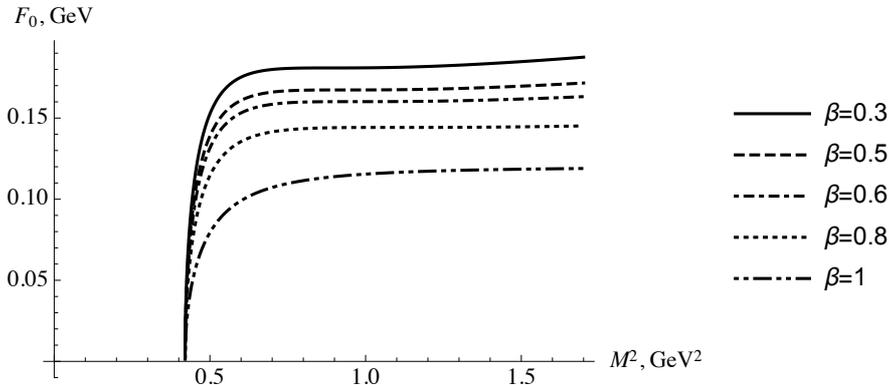}
    \caption{\small The constant $F_\phi$ on the Borel plane at different values of $\beta$.}
\end{figure}

The same analysis can be carried out for the $\omega$ and $\phi$ mesons.
The case of $\omega$ is numerically almost identical to the $\rho$ one.
In the $\phi$ channel, we take $m_s=130$~MeV for the mass of strange quark
at 1~GeV~\cite{pdg} and the same value of slope $a$~\cite{prd1,prd2,prd3,prd4}. After this substitution
(and account for the $6m_s^2/Q^2$ correction to the bare quark loop which
was neglected in the $\rho$ channel), the values of $m_\phi(M^2)$ and $F_\phi(M^2)$
are shown on Fig.~3 and Fig.~4 respectively.
It is seen that the predicted value of $\beta$ seems to be close to that
predicted for $\rho$ mesons.

We should remark, however, that the issue of large-$N_c$ masses is not well settled
and the phenomenology of linear radial trajectories can vary from paper to paper.
For this reason we would provide also a more conservative estimate for the rate of
decoupling: $\beta\approx0.3-0.8$ as is shown in Figs.~1 and~3. We believe that this
broader interval absorbs other theoretical uncertainties present in our analysis.
First of all, we used strictly linear ansatz~\eqref{6} for the mass spectrum.
The large-$N_c$ QCD sum rules with non-linear corrections to the ansatz~\eqref{6}
were considered in many
papers~\cite{duality1_1,duality1_2,duality1_3,sr0_1,sr0_2,sr0_3,sr0_4,sr0_5,sr0_6}. The experience of those models shows
that the impact of non-linear corrections competes with the accuracy of the
large-$N_c$ approximation itself. Such corrections can be added to our analysis
as well but at the price of introducing new parameters. We think that this would not
make our results substantially more convincing. Another issue is the influence of
approximating the sum over hadron states in Eq.~\eqref{3} by a infinite sequence
of stable one-meson poles. What if we relax in a reasonable manner these tight bounds?
One usually relates the caused uncertainties with violations of local quark-hadron
duality. There is no theory of such violations, only models (see, e.g.,
Refs.~\cite{sr_sh1,duality1_1,duality1_2,duality1_3}).
The violations might become important at finite $N_c$. Within the proposed models, the residues
usually acquire additional factor $F^2\rightarrow F^2/(1+K/N_c)$, with constant $K>0$. Since
we dealt with fractions of residues, this extra factors are not relevant for our analysis.
Also the spectral density acquires additional contribution, up to a normalization
factor the spectral function looks like~\cite{sr_sh1}
\begin{equation}
\rho(q^2)=\frac{1}{\pi}\text{Im}\,\Pi(q^2)\rightarrow 1+ \text{power corr.}+2e^{-Aq^2/N_c}\cos{(Bq^2)},
\end{equation}
where $A,B>0$ are some constants. The last term is present even in the limit of infinite $N_c$,
but it oscillates and therefore is not seen in the OPE in Euclidean space. It becomes
exponentially damped at finite $N_c$ so it should affect only slightly the physical observables
in the Minkowski space at $N_c=3$. The quantitative impact depends on a concrete observable.
For instance, the overall theoretical uncertainty caused by duality violations for the
cross section of hadronic $\tau$-decays was estimated in Ref.~\cite{sr_sh1} on the level
of 3\%. We recall that the expected theoretical uncertainty from the large-$N_c$ approximation
is on the level of 10-20\%. This should definitely absorb the uncertainties coming from duality
violations.

\section{Concluding discussions}

We demonstrated that the exponential decrease of couplings of light vector mesons to the $e^+e^-$ annihilation
as a function of radial number $n$ is consistent with the QCD spectral sum rules and
estimated numerically the rate of decrease. This rate turned out to be close (perhaps equal)
to that in the heavy vector mesons. The physical meaning of our result is that the radially excited
neutral vector mesons seem to decouple from the $e^+e^-$ annihilation exponentially fast with $n$
and the mechanism of this decoupling is likely universal for the light and heavy mesons.

The highly excited radial states of light vector mesons are practically not seen
(roughly starting from the second radial excitation) in the $e^+e^-$
annihilation into hadrons and it became standard to attribute
this to rapidly growing and overlapping decay widths as is modeled in the ansatz~\eqref{10}.
The representation in form of a sum of Breit-Wigner peaks is known to contradict to analytical
properties of amplitudes but we draw attention to another problem: Even "corrected" versions of such a
representation (see Ref.~\cite{sr_sh1}) predict a much slower decrease of the height of resonance peaks in the
cross-section of $e^+e^-$ annihilation to hadrons.  Our calculation suggests
that the dominant contribution to this "melting" of excited states in the perturbative continuum
may arise from the proposed effect of exponential decoupling.

Our conclusion does not necessary contradict to the large-$N_c$ prediction~\eqref{7}.
We considered a finite number of states in QCD sum rules, i.e. finite $N_c$, and the
replacement by infinite number of states in the sum was just a technical approximation which was
justified owing to a fast decrease of residues. One can easily imagine a situation when both
pictures are mutually consistent. For instance, the residues $F_n^2$ may contain an $N_c$-dependent
factor $(e^{-2/N_c})^n$ which is equal to 1 in the limit $N_c\rightarrow\infty$ (thus leading
to the prediction~\eqref{7}) and to $(e^{-2/3})^n\approx0.5^n$ in the real world as we obtained.
In this scenario, one cannot neglect the $1/N_c$ corrections to e/m couplings of radially
excited states since, in contrast to meson masses, this would lead to a dramatic disagreement
with the hadron phenomenology.

The most plausible origin of decoupling under consideration is the well known fact that
hadrons are very complicated extended objects. It means that hadron interaction with vector
current must include an e/m formfactor which may differ significantly in the large-$N_c$
limit and at finite $N_c=3$. On a qualitative level, the quarks inside excited states are much
more energetic and, consequently, on average more separated in space. This should lead to decoupling from
interaction with strictly local vector current since the couplings $F_n$ are related with the
wave function of quark-antiquark pair at zero space separation, $F_n\sim\Psi_n(0)$. We have a freedom to ascribe the effect of
non-trivial formfactor either to definition of e/m couplings $F_n$ or to definition of $e^+e^-$
decay width~\eqref{8} --- its r.h.s. should be then multiplied by a factor $\mathcal{F}_n$ that
reflects the impact of formfactor at different $n$,
\begin{equation}
\label{24}
\Gamma_{\rho_n\rightarrow e^+e^-}=\frac{4\pi\alpha^2F_n^2}{3m_n}\mathcal{F}_n.
\end{equation}
Within the former definition, we obtain
decreasing with $n$ couplings $F_n$, as we did in the present work. Within the latter one,
the couplings $F_n$ in~\eqref{24} are constant, $F_n=F_0$, in accord with the large-$N_c$ result~\eqref{7},
but predictions for $\Gamma_{\rho_n\rightarrow e^+e^-}$ must be calculated with the factor $\mathcal{F}_n$
taken into account. We may reformulate the result of our analysis as a phenomenological
derivation of this factor, $\mathcal{F}_n\approx e^{-\frac23n}$.

The decoupling scheme proposed in the present work can have various applications. For instance,
it would be interesting to employ it in the large-$N_c$ calculations
of electromagnetic pion (see, e.g., Ref.~\cite{sr_sh2,formfactor1,formfactor2}
and references therein) and proton~\cite{hernandez} fomrfactors.

\newpage{\pagestyle{empty}\cleardoublepage}

\end{document}